\documentclass[12pt, preprint,pre] {revtex4}

\usepackage{graphics}
\usepackage{verbatim}
\usepackage{amsmath}

\usepackage[tight,hang]{subfigure}

\newcommand{\lj}{Lennard-Jones}
\newcommand{\het}{heterogeneous}
\newcommand{\nm}{\phi} 
\newcommand{\order}{\phi}
\newcommand{\nn}{n_{i}} 

\newcommand{\nc}{n_{\rm c}}
\newcommand{\nb}{n_{\rm b}} 

\newcommand{\ac}{a_{\rm s}}
\newcommand{\drop}{nuc\-lea\-ting drop\-let}
\newcommand{\nuc}{nuc\-lea\-tion}
\newcommand{\ps}{pseudo\-spinodal}
\newcommand{\s}{spinodal}

\newcommand{\rthres}{r^*}
\newcommand{\kb}{k_{\rm B}}
\newcommand{\cij}{c_{ij}}	 
\newcommand{\cii}{c_{ii}} 	

\newcommand{\tc}{T_{\rm m}}	
\newcommand{\e}{\epsilon}
\begin{document}
\title{Homogeneous and heterogeneous nucleation of Lennard-Jones liquids}
\author{Hui Wang}
\affiliation{Department of Physics, Clark University, Worcester, MA 01610}
\author{Harvey Gould}
\affiliation{Department of Physics, Clark University, Worcester, MA 01610}
\author{W.\ Klein}
\affiliation{Department of Physics, Boston University, Boston, MA 02215}


\begin{abstract}
The homogeneous and heterogeneous nucleation of a Lennard-Jones liquid is investigated using the umbrella sampling method. The free energy cost of forming a nucleating droplet is determined as a function of the quench depth, and the saddle point nature of the droplets is verified using an intervention technique. The structure and symmetry of the nucleating droplets is found for a range of temperatures. We find that for deep quenches the nucleating droplets become more anisotropic and diffuse with no well defined core or surface. The environment of the nucleating droplets form randomly stacked hexagonal planes. This behavior is consistent with a spinodal nucleation interpretation. We also find that the free energy barrier for heterogeneous nucleation is a minimum when the lattice spacing of the impurity equals the lattice spacing of the equilibrium crystalline phase. If the lattice spacing of the impurity is different, the crystal grows into the bulk instead of wetting the impurity.
\end{abstract}

\maketitle

\section{Introduction}
Although nucleation from a supercooled liquid has been the subject of extensive simulations~\cite{YangPRL88,SwopePRB90, MatsumotoNature02, LeyssaleJCP05,YangJCP90,Cherne2004}, theory~\cite{AlexanderPRL78,KleinPRL86, KleinPRE01,OxtobyJCP96}, and experiments~\cite{SchofieldScience01, BodartPRB03}, the nature of the \drop\ in supercooled liquids is not well understood, especially for deep quenches.
For shallow quenches (near coexistence), classical nucleation theory applies. For deeper quenches, nucleation is affected by the proximity to the liquid-solid \s\ for systems with long-range interactions~\cite{KleinPRL86}. The spinodal represents the limit of stability of the metastable liquid and is well defined only in the limit of an infinite interaction range. However, spinodal-like effects have been found for deep quenches in systems with intermediate and short-range interactions~\cite{HeermannPRL83, YangPRL88, TruduPRL06}. Spinodal nucleation theory predicts that the decrease of the surface tension of the droplets as the spinodal is approached makes the \drop s diffuse and fractal-like. Moreover, the symmetry of the \drop s is not necessarily the same as the symmetry of the stable phase, and the symmetry of the \drop s in three dimensions is either body-centered cubic (bcc) or randomly stacked hexagonal planes. Trudu et al.~\cite{TruduPRL06} studied nucleation of a \lj\ liquid using transitional path sampling, and found a crossover from classical to spinodal-like behavior for deeper quenches. In particular, they observed that the \drop s become less compact and spherical, but did not analyze the structure and symmetry of the \drop s.

The study of \het\ nucleation, that is, nucleation that occurs on impurities, is of much practical importance because most nucleation events that occur in nature are \het. Examples include nucleation on a container wall~\cite{ToxvaerdJCP02} and nucleation of proteins in porous media~\cite{SearPNAS2006}. Many experiments~\cite{YinAM2000,HaymetPNAS2002} and simulations~\cite{FrenkelPRL2003, SearPRL2006,WarrenPRL2007, CastroPRE03} have been done to study heterogeneous nucleation. Existing theories of \het\ nucleation are mostly phenomenological and natural extensions of classical nucleation theory~\cite{WangJCP2004,LauriJCP2006, DjikaevJCP2006}. It is known that the presence of impurities can lower the free energy barrier of nucleation by as much as several orders of magnitude~\cite{FrenkelPRL2003}. The effectiveness of an impurity to decrease the nucleation barrier is determined by properties such as the shape of the impurity and the surface tension between the substrate and the metastable liquid. Page and Sear~\cite{SearPRL2006} studied heterogeneous nucleation in porous media using the Ising model and found that a pore which is approximately the size of the critical nucleus is optimal for decreasing the nucleation barrier. Heterogeneous nucleation on a structureless solid surface has also been simulated~\cite{ToxvaerdJCP02}. However, the effects of the microscopic properties of the impurities on nucleation have not been well characterized.

In this paper, we will study homogeneous and \het\ nucleation in supercooled \lj\ liquids using the umbrella sampling method. For homogeneous nucleation, we find spinodal effects for deep quenches by analyzing the structure of the \drop s. In particular, the \drop s are found to become more anisotropic and diffuse with no well defined core or surface. The droplets and their immediate environment form randomly stacked hexagonal planes, which is consistent with the spinodal nucleation picture.

To study \het\ \nuc\, a fixed impurity consisting of particles that form a hexagonal plane is added to the system. We find that the impurity whose lattice spacing is equal to the lattice spacing of the equilibrium crystalline phase is most effective in lowering the free energy barrier of nucleation. We also find that when the lattice spacing of the impurity is different than the optimal spacing, the crystal prefers to nucleate on the newly formed crystal (grow into the bulk) instead of wetting the impurity.

This paper is organized as follows. Section~\ref{sect:sim_details} describes the simulation details and the intervention technique which we use to test the saddle point nature of the nucleating droplets. Section~\ref{sect:homo_results} presents the simulation results of homogeneous nucleation, and Sec.~\ref{sect:lj_hetero} discusses our results on heterogeneous nucleation.

\section{\label{sect:sim_details}Simulation details}
The three-dimensional system of interest consists of $N = 4000$ particles with density $\rho = 0.95$ interacting via the Lennard-Jones potential. Periodic boundary conditions are used. We adopt dimensionless units so that lengths and energies are given in terms of the Lennard-Jones parameters $\sigma$ and $\epsilon$. We first prepared a liquid at $T = 1.20$, which is above the coexistence temperature ($\tc \approx 1.15$~\cite{VerletPR1969}), by melting a perfect fcc crystal; this simulation is done using the Metropolis algorithm at constant volume. The system is equilibrated for 50,000 Monte Carlo steps per particle (mcs) in the liquid phase before the quench. The system is then quenched by rescaling the temperature by a factor of 0.999 every 20\,mcs.

Because the probability of nucleation is very small, we used the umbrella sampling method~\cite{TorrieCPL74,TorrieJCompP77, FrenkelJCP92, FrenkelJCP04,FrenkelJCP96}. We denote $\nm$ as the order parameter, which we choose to be the number of particles in the largest (solid-like) cluster. (The definition of the solid-like clusters is discussed in Sec.~\ref{sect:cluster_analysis}.) The free energy $G(\nm)$ is calculated from the relation
\begin{equation}
\label{eqn:glnp}
G(\nm) = -\kb T\ln P(\nm),
\end{equation}
where $P(\nm)$ is the probability density of $\nm$.

In the umbrella sampling method, the system is sampled according to the total energy $\tilde{V} = V + V_{\rm b}(\nm)$, where $V$ is the original potential energy of the system and $V_{\rm b}(\nm)$ is the bias potential. The probability distribution $\tilde P(\nm)$ is sampled according to the total density operator $\tilde \rho = e^{-\beta V}e^{-\beta V_{\rm b}(\nm)} = \rho_{0} e^{-\beta V_{\rm b}(\nm)}$, which is the product of the original density operator $\rho_{0}$ and the weight function due to the bias $e^{-\beta V_{\rm b}(\nm)}$. The original distribution $P(\nm)$ can be determined by
\begin{equation}
P(\nm) = \tilde P(\nm)e^{\beta V_{\rm b}(\phi)}.
\end{equation}
Hence, the free energy $G(\nm)$ in Eq.~\eqref{eqn:glnp} can be calculated by
\begin{equation}
G(\nm) = -\kb T\ln \tilde P(\nm) - V_{\rm b}(\nm).
\end{equation}

As in Ref.~\onlinecite{FrenkelJCP04}, the potential bias has the form
\begin{equation}
\label{eq:bias}
V_{\rm b}(\nm) = \frac{1}{2}k(\nm - \nm_{0})^{2}.
\end{equation}
The constant $k = 0.05$ determines the width of the sampling window and yields $\Delta \nm \approx 15$. We consider a sequence of values of $\nm_{0}$ starting from size 0 and increasing by ten particles, that is, $\nm_{0}=0$, 10, 20, \ldots. Because the width of the window $\Delta \nm \approx 15$, the choice of ten particles means that the sampling windows overlap. Before collecting data for the probability $\tilde P(\nm)$ for each value of $\nm_0$, the system is equilibrated for 10,000\,mcs. The values of $\nm$ are then sampled for 100,000\,mcs. To save equilibration time a configuration for the current value of $\order_0$ is used as the initial condition for the next value of $\order_0$. Because determining the size of the largest cluster is computationally expensive, we make trial moves of 5\,mcs using only the Lennard-Jones potential without the bias potential \eqref{eq:bias}, and then accept or reject these trial moves using only the bias potential in Eq.~\eqref{eq:bias}.

\subsection{\label{sect:cluster_analysis}Cluster analysis}
Unlike Ising/Potts models~\cite{ConiglioJPA80} there is no rigorous definition of clusters in a continuous particle system. Instead we are forced to rely on our intuition to identify the solid-like particles. We use the local bond-order analysis introduced by Steinhardt et al.~\cite{SteinhardtPRB83} and developed by Frenkel and co-workers~\cite{FrenkelJCP96}. We define the $(2l + 1)$ component complex vector $\overline{q}_{lm}(i)$ for particle $i$:
\begin{equation}
\overline{q}_{lm}(i) = \frac{1}{\nn}\sum_{j = 1}^{\nn}Y_{lm}(\hat{r}_{ij}),
\end{equation}
where the sum is over the $\nn$ nearest neighbors of particle $i$ and $Y_{lm}(\hat{r}_{ij})$ is the
spherical harmonic as a function of the unit direction vector $\hat{r}_{ij}$ between particle $i$ and its $\nn$ neighbors. The nearest neighbors of a given particle are defined to be within the distance 1.4, which corresponds to the position of the first minimum of the radial distribution function $g(r)$ of the crystalline phase at the same density and temperature.

It has been shown that $l = 6$ is a good choice for characterizing the structures of crystals~\cite{FrenkelJCP92}. The rotational invariants $w_{4}(i)$, $w_{6}(i)$, $q_{4}(i)$, and $q_{6}(i)$ are defined as
\begin{align}
q_{l}(i) &= \Big(\frac{4\pi}{2l + 1}\Big)\!\sum_{m = -l}^{l}| \overline{q}_{lm}(i)|^{2})^{1/2},\\
\noalign{\noindent and}
w_{l}(i) &= \frac{\overline{w}_l(i)}{(\sum_{m = -l}^{l} |\overline{q}_{lm}(i)|^{2})^{3/2}},\\
\noalign{\noindent with}
\overline{w}_l½(i) & = \sum_{m_{1} + m_{2} + m_{3} = 0}
\begin{pmatrix} 
l & l & l\\ 
m_{1} & m_{2} & m_{3} 
\end{pmatrix}
\overline{q}_{lm_1}(i) \overline{q}_{lm_{2}}(i) \overline{q}_{lm_3}(i). \label{7}
\end{align}
The quantity in brackets in Eq.~\eqref{7}
is the Wigner's $3j$ symbol.

To define a solid-like particle, we first introduce the normalized quantity $\tilde{q}_{lm}(i)$
as
\begin{equation}
\tilde{q}_{lm}(i) = \frac{\overline{q}_{lm}(i)}{[\sum_{m = -l}^{l}| \overline{q}_{lm}(i)|^2]^{1/2}},
\end{equation}
and form the dot product
\begin{equation}
\cij = \sum_{m = -6}^{6}\tilde{q}_{6m}(i)\tilde{q}_{6m}(j)^*,
\end{equation}
where $^*$ indicates the complex conjugate. The dot product $\cii = 1$ by construction. Particles $i$ and $j$ are said to be {\it coherent} if the real part of the dot product $\cij$ is greater than 0.5. A particle is considered to be solid-like if its number of coherent neighbors is greater than or equal to $\nb$. We will chose $\nb = 11$ for reasons that we will discuss in Sec.~\ref{sect:lj_intervention_method}. After finding all the solid-like particles, we identify the clusters using the criterion that any two solid-like particles that are nearest neighbors belong to the same cluster.

As discussed in Ref.~\onlinecite{FrenkelJCP96}, the distribution of the invariants $q_{4}(i)$, $w_{6}(i)$, and $q_{6}(i)$ can be used to characterize the symmetry of a group of particles, for example, a cluster. For a given group of particles we first determine the histogram of each invariant. The three histograms are then rescaled so that they do not overlap and are combined to form the histogram $H$. The histogram $H$ corresponding to a group of particles is decomposed in terms of the histograms of the invariants corresponding to each symmetry of interest, namely,
\begin{equation}
H = f_{\rm fcc}H_{\rm fcc} + f_{\rm bcc}H_{\rm bcc} + f_{\rm hcp}H_{\rm hcp}.
\end{equation}
For example, $H_{\rm fcc}$ is determined for a system with fcc symmetry in which a certain amount of randomness has been introduced~\cite{randomness}. The coefficients corresponding to each symmetry, $f_{\rm fcc}, f_{\rm bcc}$, and $f_{\rm hcp}$, are found by minimizing the quantity
\begin{equation}
(H - f_{\rm fcc}H_{\rm fcc} - f_{\rm bcc}H_{\rm bcc} - f_{\rm hcp}H_{\rm hcp})^{2},
\end{equation}
with the constraints that $f_{\rm fcc}, f_{\rm bcc}, f_{\rm hcp} > 0$ and $f_{\rm fcc} + f_{\rm bcc} + f_{\rm hcp} = 1$. The coefficients $f_{\rm fcc}, f_{\rm bcc}$ and $f_{\rm hcp}$ are indications of the composition of each symmetry associated with the group particles.

\begin{table}[t] 
\begin{center}
\begin{tabular}{|c|c|c|c|}
\hline
T & fraction &T &fraction\\
\hline
0.75 & 0.65 & 0.60 & 0.40\\
\hline
0.70 & 0.60 &0.55 & 0.45\\
\hline
0.65 & 0.35 &0.53 & 0.55\\
\hline
\end{tabular}
\end{center}
\vspace{-0.25cm}
\caption[Fraction of the copies of the \drop s that grow after intervention]{\label{tab:growth_of_intervention}Fraction of the copies of the \drop s that grow after intervention (out of 20). These results indicate that the clusters that are generated at the maximum of the free energy correspond to \drop s.}
\end{table}

\subsection{\label{sect:lj_intervention_method}The intervention method}
Because the cluster analysis involves parameters such as the minimum number of coherent neighbors, the computed size of the clusters depends somewhat on the values chosen for the parameters. To determine if our choices are consistent, we use the fact that the \drop s correspond to the maximum of the free energy barrier and should be saddle point objects; that is, the \drop s should grow or shrink with approximately 50\% probability under a small perturbation. In practice, the parameters in the cluster analysis are first chosen and the umbrella sampling procedure is performed. Then the intervention method is used to test if intervention causes the clusters corresponding to the maximum of the free energy to grow or shrink with $\approx 50$\% probability. If these clusters do not, the parameters in the cluster analysis are modified until approximately 50\% growth probability is achieved for the modified clusters at the new free energy maximum.

\begin{figure}[t] 
\begin{center}
\scalebox{0.9}{\includegraphics{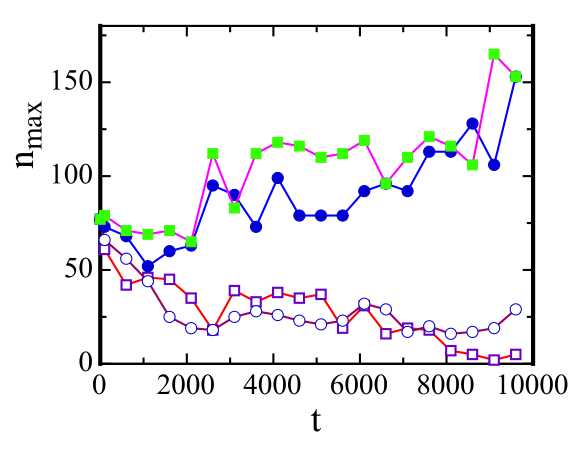}}
\vspace{-0.5cm}
\caption[Evolution of the size of the largest cluster after intervention]{(Color online) The size of the largest cluster as a function of the time after intervention for four trials. After $t_{\rm obs} = 10000$\,mcs, the largest droplet in each trial is determined to grow or shrink based on its size and the position of its center of mass compared to the largest cluster in the original configuration. The size of the \drop\ in this case is $\approx 80$ particles (for $T = 0.65$).}
\label{fig:t_65_release_121_intervention_growth_curve}
\vspace{-0.5cm}
\end{center}
\end{figure}

The intervention method we adopt is similar to what has been used to study nucleation in the Ising model~\cite{MonetteJSP92}. To test if a cluster is a \drop, we stop the simulation and make many copies of the system. Each copy is restarted with a different random number seed without the potential bias. We then determine if the largest cluster in each copy grows at approximately the same place at approximately the same time as the original. After the time $t_{\rm obs}$, both the size and location of the cluster are examined. If the size of the cluster is larger than its original size and the center of mass is within a distance $\rthres$ from the cluster in the original configuration, the cluster is said to grow. The role of $\rthres$ is to ensure that the cluster is the same as the one in the original configuration. We choose $t_{\rm obs} = 10000$\,mcs. The distance $\rthres$ should correspond to the size of the cluster; typically 50\% of the linear spatial extent of the cluster is sufficient to decide whether the cluster kept its identity during $t_{\rm obs}$. Because intervention is very time consuming, we made 20 copies of each configuration. The evolution of the size of the largest cluster after intervention for four trials is plotted in Fig.~\ref{fig:t_65_release_121_intervention_growth_curve} for $T = 0.65$. Note that the largest cluster has grown in some trials and shrunk in others. By determining the frequency of the successful trials we can estimate the probability of growth of a cluster. The fraction of trials (out of 20 trials) for which the cluster grows is listed in Table~\ref{tab:growth_of_intervention} for different temperatures. We found that setting $\nc = 11$ gives consistent results.

\section{Homogeneous nucleation}
\label{sect:homo_results}
\begin{figure}[t] 
\begin{center}
\subfigure[\ The Gibbs free energy.]{\scalebox{0.8}{\label{fig:gibbs_en_barrier_t55}\includegraphics{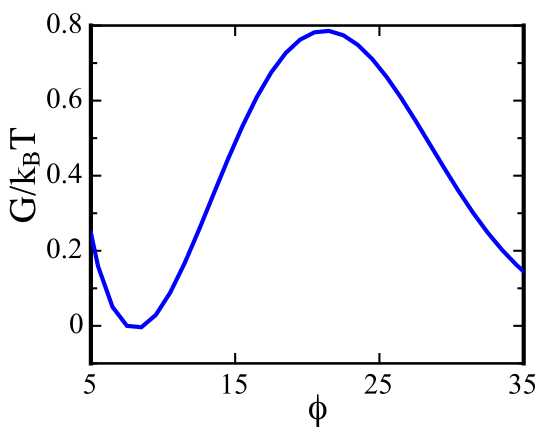}}}
\subfigure[\ The nucleation barrier.]{\scalebox{0.8}{\label{fig:gibbs_en_barrier_vs_t}\includegraphics{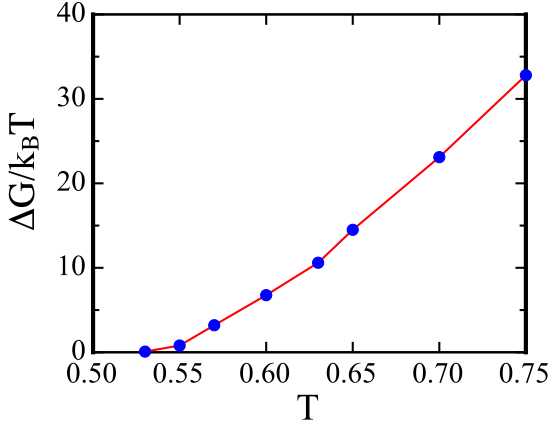}}}
\caption[The free energy landscape and its temperature dependence]{\label{fig:gibbs_en_barrier} (a) The Gibbs free energy as a function of the order parameter $\nm$, the size of the largest cluster in the system, at $T = 0.55$. The minimum of $G$ corresponds to the metastable supercooled liquid phase, and the maximum corresponds to the \drop. The free energy minimum is at $\nm \approx 7$, showing that there are some solid-like particles in the supercooled liquid state. (b) The nucleation barrier as a function of the temperature. The barrier vanishes at $T \approx 0.53$.}
\vspace{-0.25cm}
\end{center}
\end{figure}

We determined $G(\nm)$ at several temperatures between $T = 0.75$ and $T = 0.53$. As expected, $G(\nm)$ exhibits a minimum and a maximum (see Fig.~\ref{fig:gibbs_en_barrier_t55}). The minimum of $G$ corresponds to the metastable supercooled liquid, where solid-like particles are present due to thermal fluctuations. The maximum occurs at the size of the \drop\ (recall that we have chosen the order parameter to be the size of the largest cluster). The free energy difference of the maximum and the minimum can be interpreted the free energy barrier $\Delta G$ for nucleation~\cite{FrenkelJCP04}. We find that the nucleation barrier decreases with temperature (see Fig.~\ref{fig:gibbs_en_barrier_vs_t}) and vanishes at $T \approx 0.53$. This vanishing of the free energy barrier raises questions on whether or not it corresponds to a spinodal.

The spinodal is usually defined as the sharp boundary between the metastable and unstable states~\cite{BinderPRA84}. In particular, the \s\ is a thermodynamic transition that acts as a line of critical points. (In Ising models the \s\ corresponds to a divergent isothermal susceptibility~\cite{HeermannPRL82}.) We will refer to this interpretation as the classical spinodal. The spinodal as defined in this way is present only in mean-field systems such as those with infinite range interactions. Systems with large but finite range interactions can exhibit \ps\ effects~\cite{KleinPRE07}. To test if $T = 0.53$ corresponds to the classical spinodal we also simulated the system at this temperature by a standard Metropolis algorithm. By tracking the size of the largest cluster we found that the lifetime of the metastable state is in the range $[1\times 10^{5},\, 6\times 10^{5}]$\,mcs, which implies that the free energy barrier to nucleation has not vanished at $T=0.53$. Hence the vanishing of the free energy barrier found by umbrella sampling does not necessarily correspond to a classical spinodal. Moreover, we found (see Table~\ref{tab:growth_of_intervention}) that the droplets found by umbrella sampling appear to be saddle point objects as determined by the intervention method (without the bias potential). A possible explanation is that the interpretation of the umbrella sampling results for $P(\nm)$ assumes that clusters whose size are comparable to the \drop\ are rare~\cite{FrenkelJCP04}. This assumption is not applicable for $T \approx 0.53$ because there is typically more than one large cluster of comparable size in the system. We will investigate this assumption and other possible explanations in future work.

\begin{figure}[htbp]
\setlength{\subfigcapskip}{0pt}
\setlength{\subfigtopskip}{0in}
\begin{center}
\subfigure[~$T = 0.75$.]{\scalebox{0.45}{\includegraphics{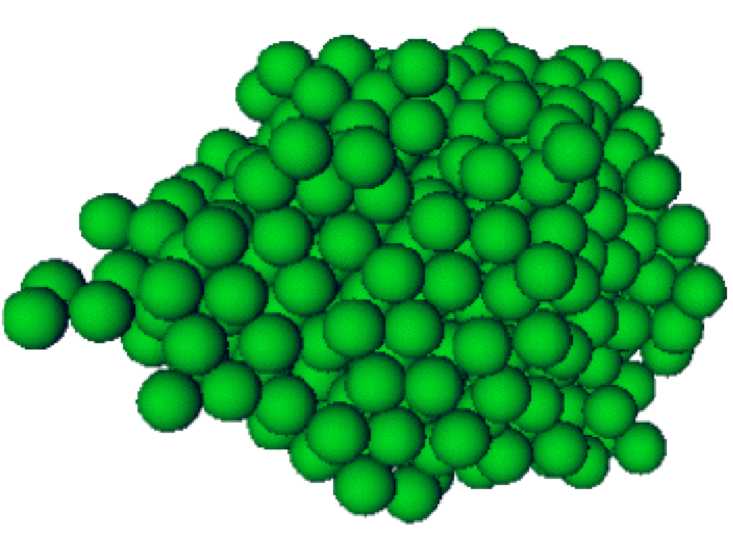}}}\hspace{2cm}
\subfigure[~$T = 0.65$.]{\scalebox{0.45}{\includegraphics{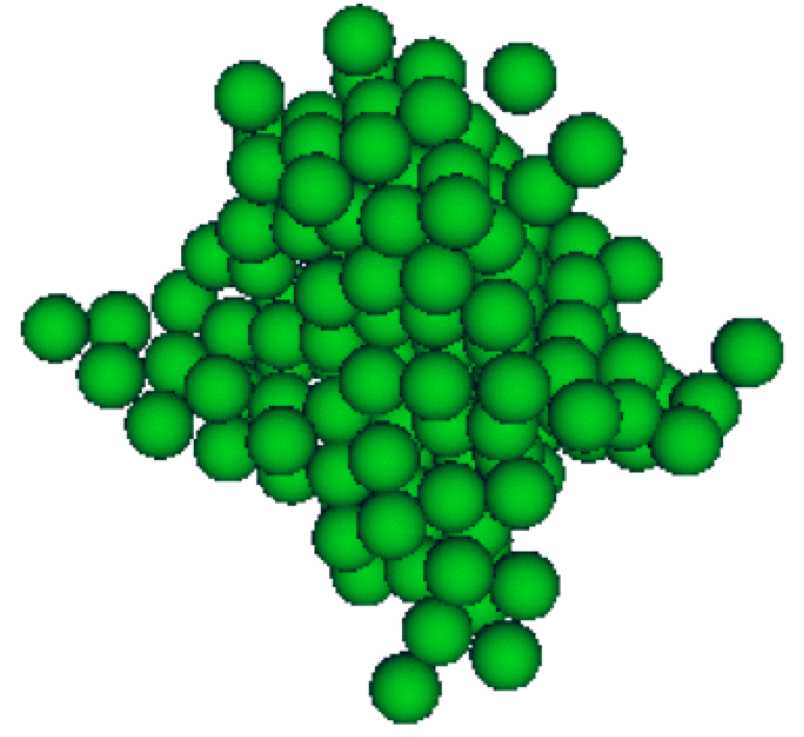}}}
\subfigure[~$T = 0.60$.]{\scalebox{0.45}{\includegraphics{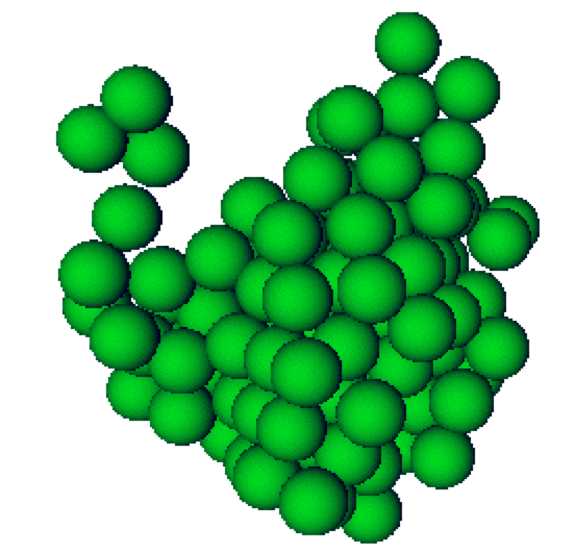}}}\hspace{2cm}
\subfigure[~$T = 0.55$.]{\scalebox{0.45}{\includegraphics{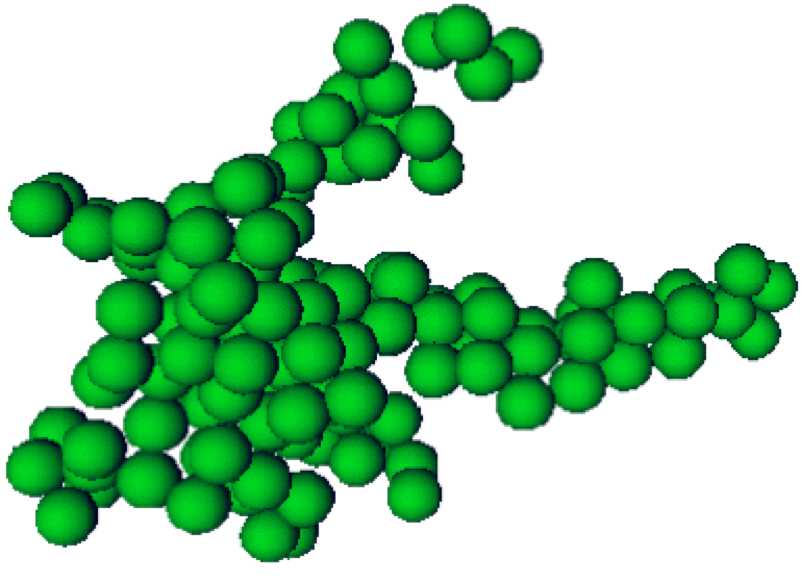}}}
\caption[Snapshots of the \drop\ at various temperatures]{\label{fig:nuclei_snap}(Color online) Snapshots of the \drop\ at different temperatures.}
\vspace{-0.25cm}
\end{center}
\end{figure}

Figure~\ref{fig:nuclei_snap} shows snapshots of the \drop\ at different temperatures. Note that the droplets are compact for moderate supercooling and become more diffuse for deeper quenches. This qualitative observation is consistent with Ref.~\onlinecite{TruduPRL06}. We will analyze the structure of the nuclei in the following.

\subsection{The structure of the nucleating droplets}
To measure the compactness of a nucleating droplet, we determine its density profile $\rho(r)$, which is defined in terms of the mean number of particles $N(r)$ in the spherical shell between $r$ and $r + dr$
\begin{equation}
N(r) = \rho(r)4\pi r^{2}dr.
\end{equation}
Here $r$ is measured from the center of mass of a \drop.

\begin{figure}[htbp]
\begin{center}
\scalebox{1.0}{\label{fig:nuclei_density_profile}\includegraphics{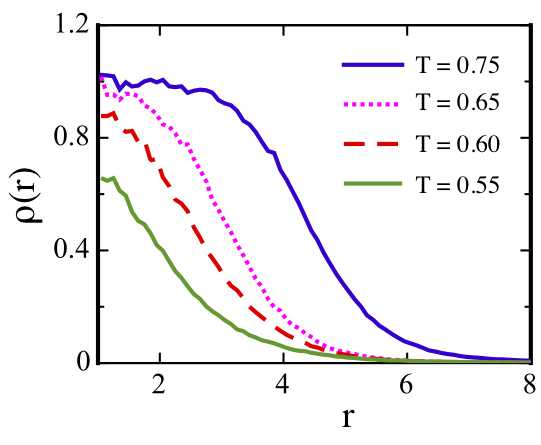}}
\vspace{-0.5cm}
\caption[The density profile $\rho(r)$ of the nucleating droplets]{\label{fig:lj_nuclei_density_profile_ab} The density profile $\rho(r)$ of the nucleating droplets for various temperatures. At $T = 0.75$, the \drop\ has a well defined core corresponding to the plateau of $\rho(r)$. Note that the density at the core is slightly higher than 0.95, the mean density. The decrease of $\rho(r)$ for larger $r$ implies that there is a well defined interface between the core and the liquid environment. At lower temperatures, the plateau disappears and the density changes gradually from the core to the surface, suggesting that the \drop\ are more diffuse.}
\vspace{-0.25cm}
\end{center}
\end{figure}

Figure~\ref{fig:lj_nuclei_density_profile_ab} shows $\rho(r)$ averaged over 1000 \drop s at the value of $\order_0$ corresponding to the maximum of $G(\nm)$. For $T = 0.75$, $\rho(r)$ has a plateau for small $r$, meaning that the droplet has a well defined core. The decrease of $\rho(r)$ for larger $r$ indicates that there is an interface between the core and the liquid environment. At $T = 0.55$, the plateau disappears, and the density slowly decreases from the core to the surface, indicating that the \drop\ is diffuse.

\begin{figure}[htbp]
\begin{center}
\subfigure{\scalebox{0.8}{\includegraphics{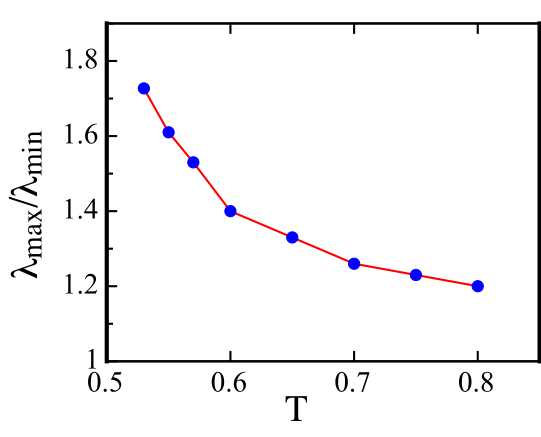}}}
\caption[Measure of anisotropy of the \drop s]{\label{fig:inertia_ratio_vs_t}The ratio of the maximum and minimum of the eigenvalues of the moment of inertia tensor of the nucleating droplets. The increase of the ratio at low temperatures indicates that the nucleating droplets become more anisotropic.}
\vspace{-0.25cm}
\end{center}
\end{figure}

To quantify the anisotropy of the nucleating droplets, we calculate the moment of inertia tensor associated with each droplet,
\begin{equation}
I_{\alpha \beta} = \sum_{i = 1}^{n}(r_{i}^{2}\delta_{\alpha \beta} - r_{i, \alpha}r_{i, \beta}),
\end{equation}
where $r_{i}^{2} = \sum_ \alpha r_{i, \alpha}r_{i, \alpha}$, $i$ labels the particles, and $\alpha$ and $\beta$ label the components of $\vec r$. The square root of the eigenvalues of $I_{\alpha \beta}$ define the principal radii of the ellipsoid characterizing the droplet. The orientation of each individual \drop\ is found to be random; that is, the long axis of the computed ellipsoid points in random directions independent of the orientations of the simulation cell. We can characterize each droplet's anisotropy by calculating the ratio of the maximum and minimum principal radii (denoted by $\lambda_{\max}$ and $\lambda_{\min}$ respectively). This ratio is one for a perfectly spherical droplet and is greater than one if the droplet is anisotropic. Figure~\ref{fig:inertia_ratio_vs_t} shows the ratio (averaged over 1000 \drop s at each temperature) as a function of the temperature. The ratio is close to one for shallow quenches, meaning that the \drop s are close to spherical. The increase of the ratio at lower temperatures indicates that the \drop s become more anisotropic. The anisotropic character of the \drop s is important in the calculation of the nucleation barrier even in classical nucleation theory~\cite{TruduPRL06}.
 
\begin{figure}[b]
\begin{center}
\subfigure[\ Mean number of particles.]{\scalebox{0.85}{\label{fig:nc_vs_t}\includegraphics{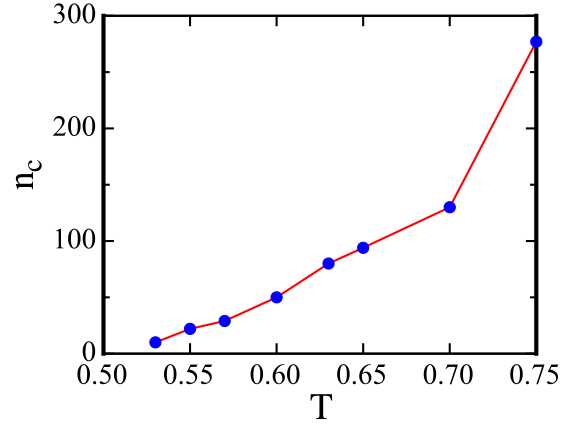}}}
\subfigure[\label{fig:majoraxis} Radius of gyration and semimajor axis.]{\scalebox{0.85}{\label{fig:radius_of_gyration}\includegraphics{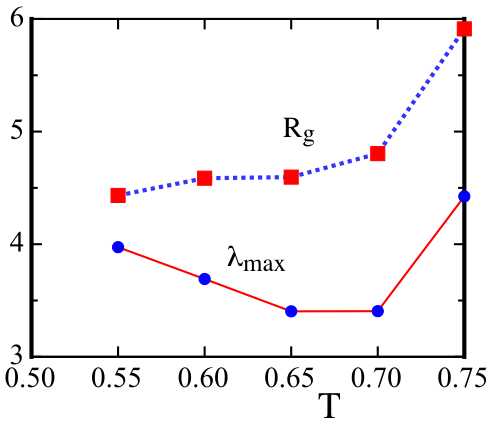}}}
\caption[The mass and radius of gyration of the \drop s as a function of temperature]{\label{fig:gibbs_nc_rg_vs_t}(a) The mean number of particles in the \nuc\ as a function of $T$. (b) The radius of gyration $R_{g}$ and mean semimajor axis $\lambda_{\max}$ of the \drop s. The data is averaged over 1000 independent configurations. Note that the $\lambda_{\max}$ increases as $T$ is decreased below $T \approx 0.65$.}
\vspace{-0.2cm}
\end{center}
\end{figure}

Figure~\ref{fig:gibbs_nc_rg_vs_t} shows that the number of particles $\nc$ in the \drop\ and the radius of gyration $R_{g}$ decrease as the temperature is decreased. Both quantities are predicted to first decrease as the temperature is lowered from coexistence and then begin to increase as the spinodal is approached~\cite{UngerPRB84}. In particular, if the \s\ interpretation is applicable, simple scaling arguments~\cite{gyration} suggest that $\nc$ (in three dimensions) and $R_g$ (in all dimensions) should increase as the spinodal is approached if the system is sufficiently close to the \s\ so that the core of the \drop\ has disappeared. The fact that $\nc$ and $R_{g}$ do not increase rapidly at lower temperatures in our simulations might be due to the nonexistence of \s\ effects and/or the underestimate of $\nc$ and $R_{g}$ near the \s\ due to our ad hoc definition of solid-like particles. A more likely explanation is that because the \drop s near the \s\ are anisotropic and effectively two-dimensional, the simple scaling arguments do not apply. In Fig.~\ref{fig:majoraxis} we plot the semimajor axis $\lambda_{\max}$ of the \drop s. Note that $\lambda_{\max}$ does show the expected behavior. In addition, the scaling arguments~\cite{gyration} in two dimensions suggest that $\nc$ is either a constant or is logarithmically divergent as the spinodal is approached. More work is needed to understand the temperature-dependence of $\nc$, $R_g$, and $\lambda_{\max}$ in the intermediate region where neither the classical nor spinodal picture is applicable.

\subsection{Symmetry of the nucleating droplets}

\begin{table}[b] 
\begin{center}
\begin{tabular}{|c|c|c|c|}
\hline
T & $f_{\rm fcc}$& $f_{\rm hcp}$& $f_{\rm bcc}$\\
\hline
0.75& 0.62 &0.38&0.00\\
\hline
0.70& 0.50 & 0.48&0.02\\
\hline
0.65& 0.46 &	0.52&0.02\\
\hline
0.60 &0.28& 0.66& 0.06\\
\hline
0.55 &0.26& 0.68& 0.06\\
\hline
\end{tabular}
\caption[The $f$ coefficients of the \drop s at different temperatures]{\label{tab:structuralComposition}Values of the coefficients corresponding to the symmetry of the \drop s for different temperatures.}
\end{center}
\end{table}

\begin{figure}[t]
\begin{center}
\scalebox{0.9}{\includegraphics{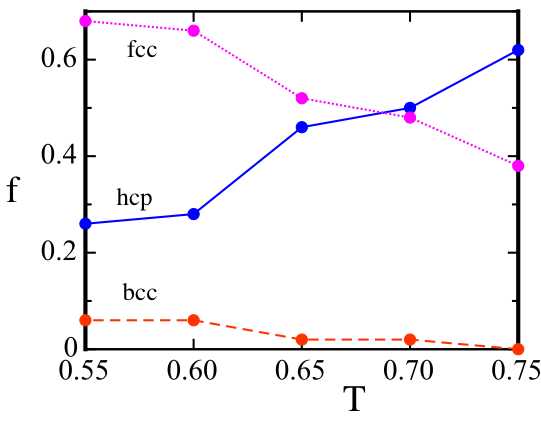}}
\vspace{-0.4cm}
\caption[The component of each symmetry of the \drop\ as a function of temperature]{\label{fig:nuclei_str_of_whole_clust_vs_T}The coefficient $f$ of each symmetry of the \drop\ as a function of temperature. The results are averaged over 1000 independent configurations. As the temperature is decreased, the hcp and bcc components increase and the fcc component decreases.}
\vspace{-0.3cm}
\end{center}
\end{figure}

The symmetry of the nucleating droplets is analyzed using the method discussed in Sec.~\ref{sect:cluster_analysis}. At each temperature, we obtain the parameters $f_{\rm fcc}$, $f_{\rm bcc}$, and $f_{\rm hcp}$ for the \drop s averaged over 1000 independent configurations. The fitting parameters are listed in Table~\ref{tab:structuralComposition} and are plotted in Fig.~\ref{fig:nuclei_str_of_whole_clust_vs_T}. As the temperature is decreased, the fcc component decreases and the bcc and 
hcp components increase. The mixture of fcc and hcp signifies the occurrence of the rhcp structure~\cite{SchofieldScience01} and is consistent with the picture of stacked hexagonal planes~\cite{KleinPRL86}.

\begin{figure}[t]
\begin{center}
\subfigure[\ $T = 0.75$.]{\scalebox{0.7}{\includegraphics{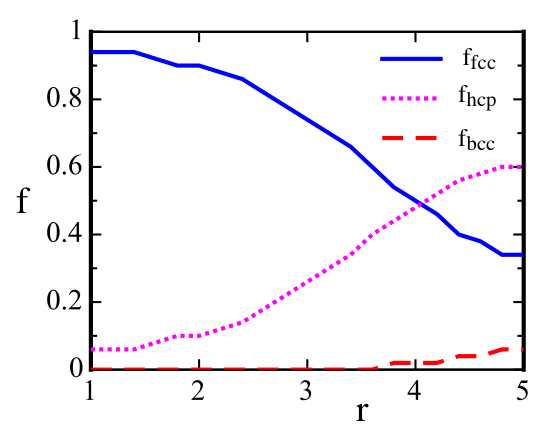}}}
\subfigure[\ $T = 0.65$.]{\scalebox{0.7}{\includegraphics{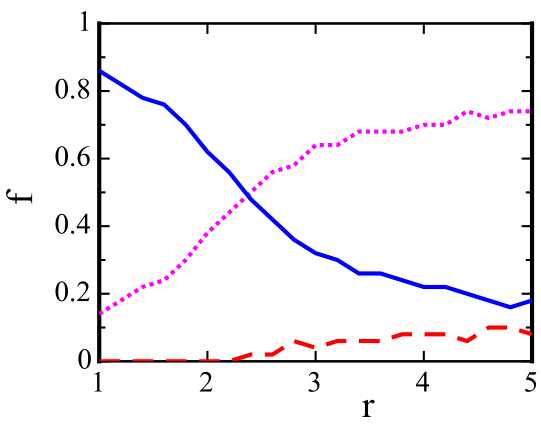}}}\\
\subfigure[\ $T = 0.60$.]{\scalebox{0.7}{\includegraphics{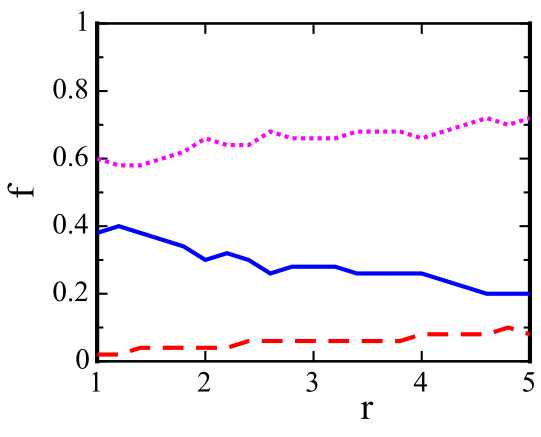}}}
\subfigure[\ $T = 0.55$.]{\scalebox{0.7}{\includegraphics{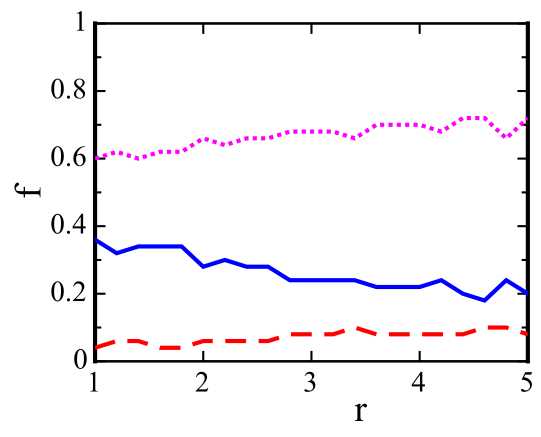}}}
\caption[The component of each symmetry as a function of $r$ for the \drop s]{\label{fig:nuclei_str}The component of each symmetry as a function of the distance $r$ from the center of mass of the \drop s at different temperatures. The results are averaged over $1000$ configurations. At $T = 0.75$, the core of the nucleus is predominately fcc, and the surface shows some bcc and liquid symmetry. At $T = 0.55$, the fcc symmetry is less dominant for small $r$, and the bcc component for small $r$ is close to its value for larger $r$.}
\vspace{-0.3cm}
\end{center}
\end{figure}

We also calculated the symmetry of the particles in a spherical shell between $r$ and $r + \Delta r$, where $r$ is measured from the center of mass of the cluster and $\Delta r$ is the thickness of the shell ($\Delta r = 0.2$). Figure~\ref{fig:nuclei_str} shows the component of each structure as a function of $r$ for various temperatures. At $T = 0.75$, the core of the
nucleating droplet is mostly fcc. Away from the center, the fcc component decreases and the bcc and liquid component increases. At
the surface, the bcc component levels off to $f_{\rm bcc} \approx 0.1$. The fact that the nucleating droplet is composed of an fcc
core and a bcc halo agrees with previous results~\cite{FrenkelJCP96}. At $T= 0.55$ the \drop\ is mainly a mixture of fcc and hcp, with a slight increase of the bcc component. The increase of bcc symmetry and decreased distinction between the bulk and the surface for deep quenches is in agreement with the spinodal \nuc\ picture~\cite{KleinPRL86}.

\begin{figure}
\setlength{\subfigcapskip}{0pt}
\setlength{\subfigtopskip}{0in}
\begin{center}
\subfigure[]{\scalebox{0.4}{\includegraphics{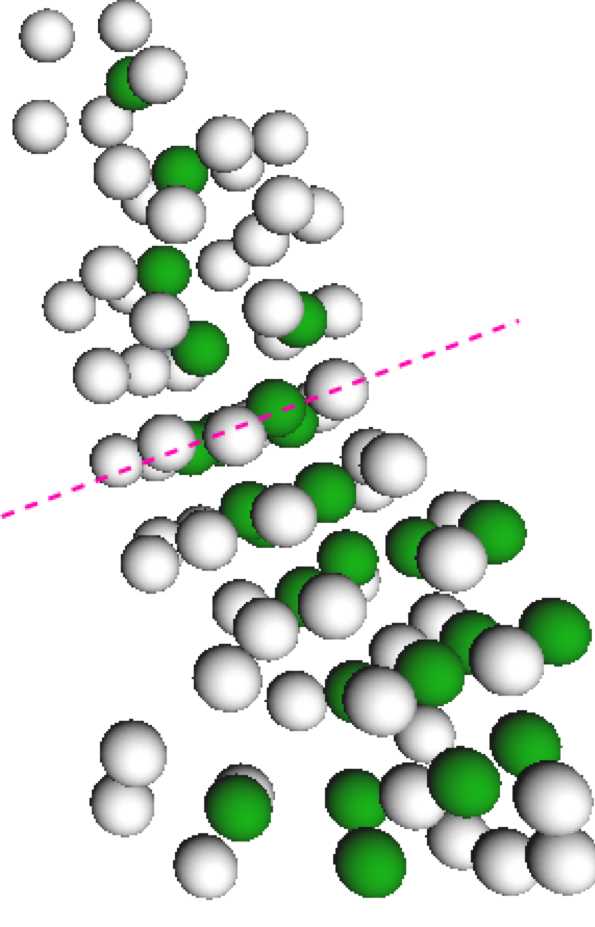}}}
\subfigure[]{\scalebox{0.6}{\includegraphics{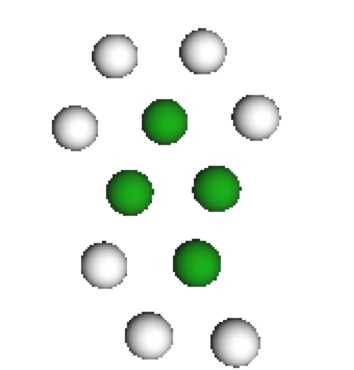}}}
\caption[A \drop\ and its local environment form stacked hexagonal planes]{\label{fig:stacked_planes1}(Color online) (a) Snapshot of a \drop\ (dark) and particles in the local environment (light) at $T = 0.55$. (b) The layer of particles corresponding to the plane indicated by the dashed line in (a) form a hexagonal structure.}
\vspace{-0.5cm}
\end{center}
\end{figure}

We also examined the structure of the particles in the \drop\ and its local environment, which consists of particles that are nearest neighbors of any particle in the droplet. Although the particles in the \drop s by themselves do not seem to form a visually identifiable structure, the \drop s and the surrounding particles in the liquid phase together form hexagonally stacked planes (see Fig.~\ref{fig:stacked_planes1}).

\section{\label{sect:lj_hetero}Heterogeneous nucleation}
\begin{figure}[t]
\begin{center}
\scalebox{0.5}{\includegraphics{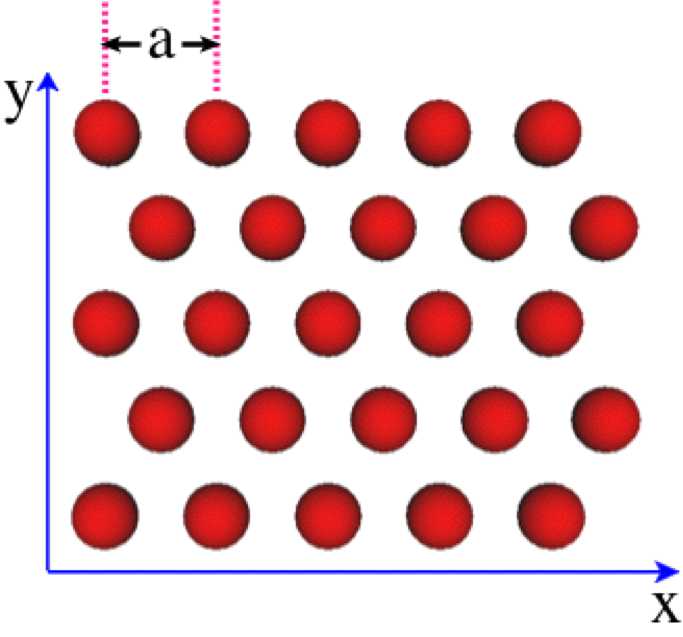}}
\vspace{-0.5cm}
\caption[Sketch of the impurity used to study heterogeneous nucleation]{\label{fig:dirt25}Sketch of the impurity used to study heterogeneous nucleation. The area of the impurity is $A = \frac{\sqrt{3}}{2}mna^{2}$, where $m$ and $n$ are the number of particles in the $x$ and $y$ directions respectively.}
\vspace{-0.25cm}
\end{center}
\end{figure}

To study \het\ nucleation, an impurity of $m\times n$ Lennard-Jones particles in a hexagonal plane is placed into the system, where $m$ and $n$ are the number of particles in the $x$ and $y$ directions (see Fig.~\ref{fig:dirt25}). The $z$ direction is perpendicular to the plane of the impurity. The positions of the particles in the impurity are fixed during the simulation. The impurity is characterized by its lattice spacing $a$ and total area $A = \frac{\sqrt{3}}{2}mna^{2}$. The efficiency of the impurity is measured by the height of the nucleation barrier $\Delta G$, which we compute as before using the umbrella sampling method.

If the system is crystallized homogeneously after a quench to $T <\tc$, the position of the first peak of the radial distribution function $g(r)$ is $\approx 1.09$. We take the value $\ac = 1.09$ as the lattice spacing of the solid phase. The temperature is quenched to $T = 0.75$, which corresponds to the region where classical nucleation applies in the absence of an impurity (see Sec.~\ref{sect:homo_results}). At $T = 0.75$ the free energy barrier of homogeneous nucleation is $\Delta G/\kb T \approx 40$ in the absence of an impurity with $\nc\approx 300$. In all of our simulations of \het\ \nuc, nucleation always occurs on the impurity if it is present.

\begin{table}[t] 
\begin{center}
\begin{tabular}{|c|c|c|}
\hline
$m \times n$ & lattice spacing $a$ & $\Delta G/\kb T$\\
\hline
$6\times6$ & 0.908 & 37\\
\hline
$5\times6$ & 0.995 & 22\\
\hline
$5\times5$ & 1.090 &3.3\\
\hline
$4\times5$ & 1.218 &20\\
\hline
$4\times4$ & 1.360&30\\
\hline
\end{tabular}
\end{center}
\caption{Values of $m$ and $n$ used to study the effect of changing the lattice spacing. The area of the impurity is fixed.
\label{tab:spacing_geometries_studied}}
\end{table}

\begin{figure}[h]
\begin{center}
\scalebox{0.75}{\includegraphics{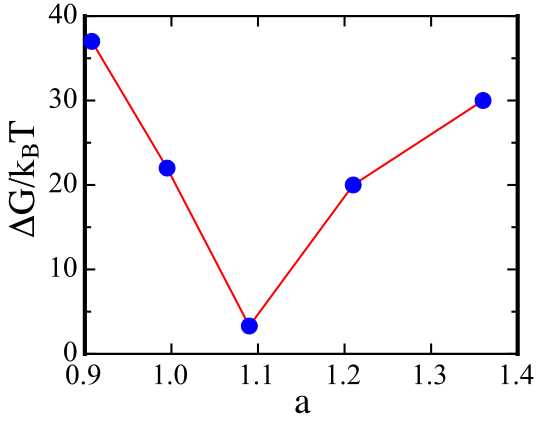}}
\caption{\label{fig:hetero_g_vs_spacing_A_25_109}The free energy barrier as a function of the lattice spacing $a$ for fixed area $A = \frac{\sqrt{3}}{2}(25\times1.09^{2})$. The nucleation barrier is a minimum at $a = \ac$.}
\vspace{-0.25cm}
\end{center}
\end{figure}

We studied the dependence of the nucleation barrier on the lattice spacing for fixed area $A$, which is chosen to be $A = \frac{\sqrt{3}}{2}(25\times 1.09^{2})$, the area of a $5\times5$ impurity with lattice spacing $\ac = 1.09$. The impurities have different values of $m$ and $n$ so that their lattice spacings are less than and greater than $\ac$ (see Table~\ref{tab:spacing_geometries_studied}). The free energy barrier is lowest at $a\approx \ac$ (see Fig.~\ref{fig:hetero_g_vs_spacing_A_25_109}), that is, an impurity is most efficient in lowering the nucleation barrier if its lattice spacing is the same as that of the crystalline phase.

\begin{figure}[htbp]
\setlength{\subfigcapskip}{0pt}
\setlength{\subfigtopskip}{0in}
\begin{center}
\subfigure[\ $a = 1.09$.]{\scalebox{0.6}{\label{fig:hetero_hex_snap_n_25_a_109}\includegraphics{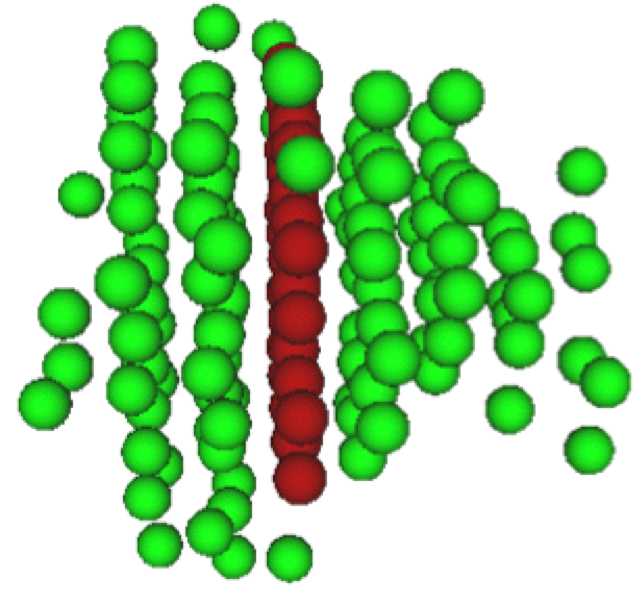}}}\hspace{1.0cm}
\subfigure[\ $a = 0.908$.]{\scalebox{0.6}{\label{fig:hetero_hex_snap_n_36_a_908}\includegraphics{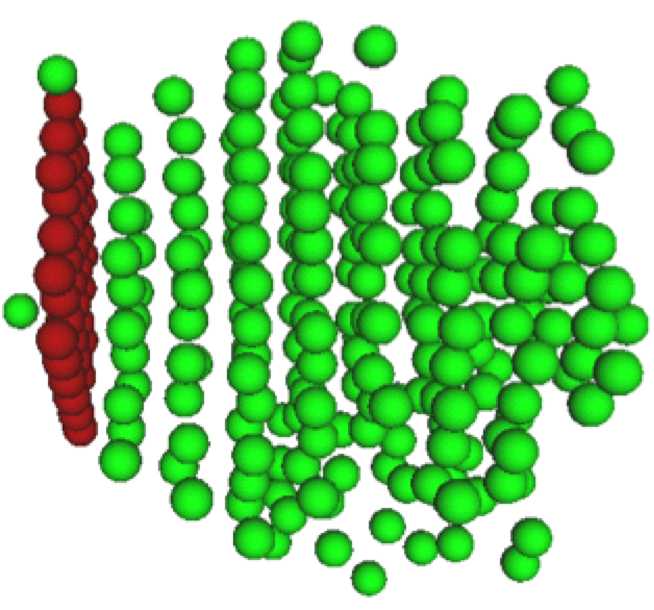}}}
\caption[Snapshots of the nucleating droplets in the presence of an impurity]{\label{fig:hetero_hex_snaps}(Color online) Snapshots of nucleating droplets in the presence of an impurity. The dark particles are the impurity; the size of the nucleating droplets is 146 and 250 particles respectively. For $a = 1.09$, the nucleating droplet grows on both sides of the impurity. For $a = 0.908$, once the impurity initiates nucleation it is preferential to add particles on the newly formed droplet than wetting the other side of the impurity.}
\vspace{-0.45cm}
\end{center}
\end{figure}

\begin{figure}[htbp]
\setlength{\subfigcapskip}{0pt}
\setlength{\subfigtopskip}{0in}
\begin{center}
\subfigure[\ Impurity and first layer of droplet.]{\scalebox{0.6}{\label{fig:hex_n_25_a_109_hexplanes_with_impurity}
\includegraphics{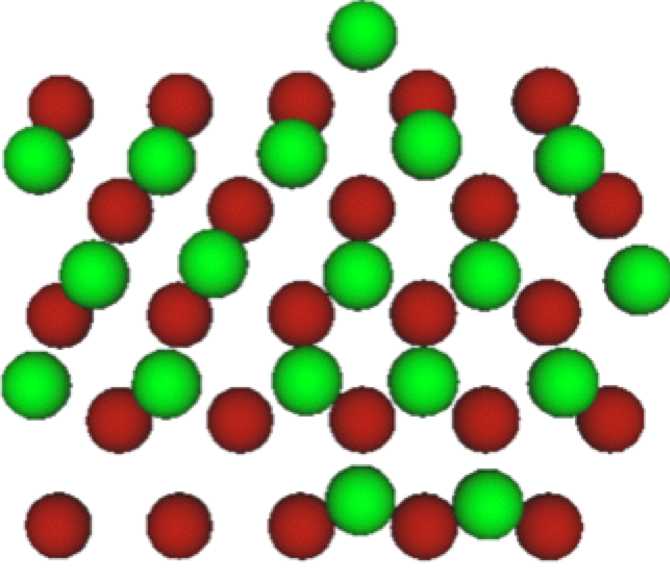}}}
\hspace{2cm}
\subfigure[\ First layer of droplet only.]{\scalebox{0.6}{\label{fig:hex_n_25_a_109_hexplanes}\includegraphics{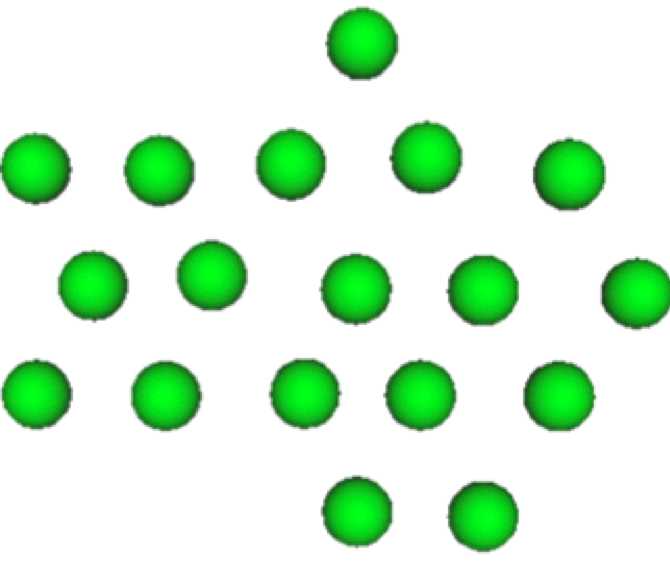}}}
\caption[Snapshots of the first layer of the \drop\ in the presence of an impurity]{\label{fig:hex_n_25_a_109_layers}(Color online) Snapshots of the first layer of the \drop\ in the presence of a $5\times5$ impurity at $a = \ac$. The layer of the nuclei lies on top of the impurity (dark particles). Within the layer the solid-like particles form a hexagonal structure. (b) Same snapshot as (a) with particles in the impurity not shown.}
\vspace{-0.25cm}
\end{center}
\end{figure}

By investigating snapshots of the \drop s we found that in general the \drop\ forms on both sides of the impurity for $a\approx \ac$ (see Fig.~\ref{fig:hetero_hex_snap_n_25_a_109}). For $a\neq\ac$, once the impurity initiates nucleation, it becomes preferential to add particles on the newly formed droplet than on the impurity, making the droplet grow into the bulk instead of wetting the other side of the impurity. In both cases, the \drop s grow on the impurity by forming layered planes, with each layer being parallel to the plane of the impurity. Inside each layer, the solid-like particles form a hexagonal structure (Fig.~\ref{fig:hex_n_25_a_109_layers}). The lattice spacing of the \drop s is approximately $\ac$ irrespective of the lattice spacing of the impurity. That is, once a nucleus is formed, its lattice spacing becomes very close to the optimal spacing $\ac$ and the droplets form on the newly formed layer rather than on the impurity.

\begin{figure}[t]
\begin{center}
\scalebox{0.8}{\includegraphics{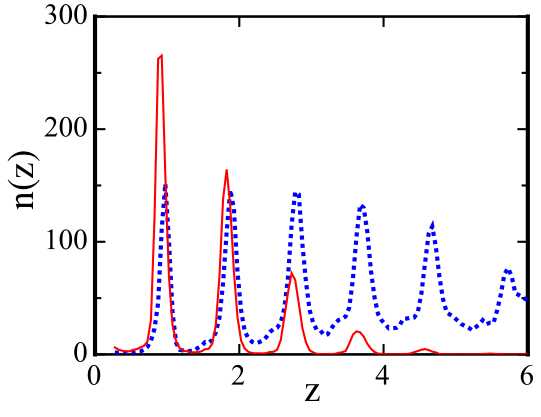}}
\vspace{-0.25cm}
\caption[Number of particles in the nuclei as a function of $z$.]{\label{fig:hetero_nz_n_36_a_908_vs_n_25_a_109}Number of particles in the nuclei as a function of $z$. The solid line is for a $5\times5$ impurity with $a = 1.09$ and the dotted line is for a $6\times6$ impurity with $a = 0.908$.}
\vspace{-0.25cm}
\end{center}
\end{figure}

A simple way to characterize how the \drop\ wets the impurity is to measure its profile. In Fig.~\ref{fig:hetero_nz_n_36_a_908_vs_n_25_a_109} we show the number of particles in the \drop\ $n(z)$ as a function of $z$, where $z$ is the distance of a particle from the plane of the impurity. The sharp peaks of $n(z)$ indicates that the nuclei form a layered structure. For $a = 1.09$ $n(z)$ gradually drops to zero as $z$ increases. In contrast, $n(z)$ is almost flat for $a = 0.908$. The protrusion of the nuclei into the bulk for $a = 0.908$ is consistent with the fact that the \drop\ wets the impurity at $a = \ac$ and grows into the bulk when $a \neq \ac$.

\begin{figure}[htbp]
\setlength{\subfigcapskip}{-10pt}
\setlength{\subfigtopskip}{0in}
\begin{center}
\subfigure[\ $a = 1.09$.]{\scalebox{0.7}{\label{fig:hetero_contour_n_25_a_109}\includegraphics{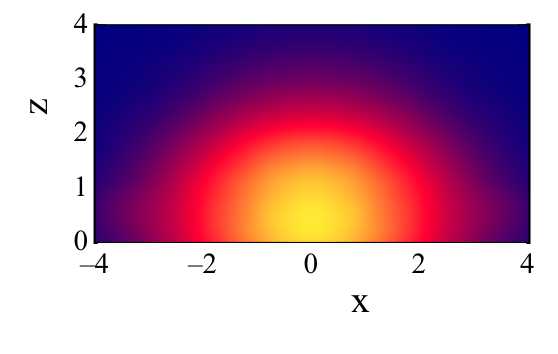}}}\hspace{1cm}
\subfigure[\ $a = 0.908$.]{\scalebox{0.7}{\label{fig:hetero_contour_n_36_a_908}\includegraphics{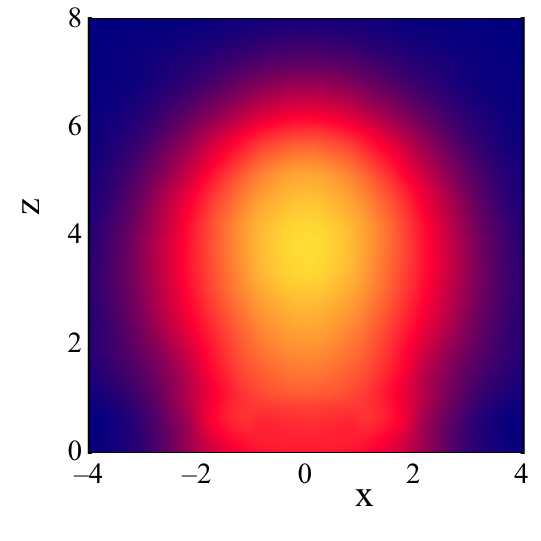}}}
\caption[Contour plots of the nucleating droplets in the presence of an impurity]{\label{fig:hetero_contour}(Color online) Contour plots of the nucleating droplets in the presence of an impurity. The impurity is at $z = -1$ and the density of the nuclei is projected onto the $x$-$z$ plane. The plots are obtained by averaging over 200 independent configurations of the nucleating droplets. (a) The lattice spacing of the impurity is $a = 1.09$, with $\nc = 150$. (b) The The lattice spacing is $a = 0.908$ with $\nc = 300$. The smaller contact angle of the impurity with $a = 1.09$ is consistent with the fact that it is preferential to nucleate on the impurity when its lattice spacing is optimal.}
\vspace{-0.25cm}
\end{center}
\end{figure}

Contour plots of the profile were obtained by projecting the density of the nuclei onto the $x$-$z$ plane. Figure~\ref{fig:hetero_contour} shows contour plots for impurities with $a = 1.09$ and $a = 0.908$ respectively. The smaller contact angle for the impurity with $a = 1.09$ is consistent with the fact that it is preferential to nucleate on the impurity when its lattice spacing is optimal. 

\section{Conclusions}
We have studied the homogeneous and heterogeneous nucleation of Lennard-Jones liquids using the umbrella sampling method. By analyzing the symmetries of the \drop s, we found that for deep quenches the nucleating droplets are more diffuse and anisotropic with no well defined core or surface; the \drop s and the corresponding liquid environment form randomly stacked hexagonal planes. These results are consistent with the spinodal nucleation picture. For \het\ nucleation, we found that the droplets grow on an impurity of hexagonal plane by layers, and the solid-like particles in each layer form a hexagonal structure. For fixed area of the impurity, the free energy barrier of nucleation is a minimum when the lattice spacing of the impurity is equal to $\ac$, the lattice spacing of the equilibrium crystalline phase. The lattice spacing of the nuclei is equal to $\ac$ even when the lattice spacing of the impurity is different than $\ac$, and it is favorable for the \drop s to grow into the bulk instead of wetting the impurity.


\begin{thebibliography}{99}
\bibitem{randomness}Because a system of Lennard-Jones particles does not have a bcc or hcp phase, we instead prepared systems with the desired symmetry and then randomly displaced each particle by an amount $(2r - 1)\delta$, where $\delta = 0.12$ and $r$ is a uniform random number between 0 and 1.
\bibitem{gyration}We can determine the temperature-dependence of the mean number of particles in the nucleating droplet $\nc$ near the \s\ from scaling arguments (see Ref.~\onlinecite{KleinPRE07}). The density of the fundamental clusters near the spinodal is $\e/G_s$, where $\e = (T-T_s)/T_s$, the Ginzburg parameter $G_s = R^d\e^{3-d/2}$, and $R$ is the effective range of the interaction. Because the \drop\ is formed by the coalescence of $G_s$ fundamental clusters~\cite{KleinPRE07}, $\nc$ scales as $\nc \sim G_s(\e/G_s)\xi^d = \e R^d \e^{-d/2} = R^d \e^{1-d/2}$, where the correlation length $\xi$ scales as $\xi \sim R\e^{-1/2}$. We conclude that $\nc$ diverges near the spinodal as $\e^{-1/2}$ in three dimensions, and remains constant (or diverges logarithmically) in two dimensions. The radius of gyration $R_{g}$ scales as $R_{g}\sim \xi \sim R \e^{-1/2}$ and diverges near the spinodal in all dimensions. Note that these scaling arguments do not predict how close the system has to be to the \s\ for the predicted behavior to be observed and assume that the \drop\ is spherical on average.
\end{thebibliography}
\end{document}